\documentclass{amsart}
\usepackage{graphicx}
\usepackage{amssymb}
\usepackage{amsmath}

\begin{document}


\title[The Clausius inequality]
{Is the Clausius inequality a consequence of the second law?}
\author{Alexey~V.~Gavrilov}

\maketitle

\begin{abstract}

We present an analysis of the foundations of the well known Clausius
inequality. It is shown that, in general, the inequality is not a
logical consequence of the Kelvin-Planck formulation of the second
law of thermodynamics. Some thought experiments demonstrating the
violation of the Clausius inequality are considered. The possibility
of experimental detection of the violation is pointed out.

\end{abstract}


\section{INTRODUCTION}

The Clausius inequality is a well known statement of classical
thermodynamics. It asserts that the integrated heat absorbed by a
system in a cyclic thermodynamic process, divided by the temperature
at which that heat is taken, is bounded from above by zero:
$$\oint\frac{\delta Q}{T}\le 0.\eqno{(1)}$$
The denominator $T$ in (1) denotes the temperature of the heat bath
from which the system takes heat $\delta Q$. In the process the
system may be in contact with several baths at different
temperatures, one at a time.

In the textbooks on thermodynamics the Clausius inequality is
usually considered an equivalent of the second law. One of the aims
of this paper is to show that the inequality (1) is not a logical
consequence of the Kelvin-Planck formulation of the second law of
thermodynamics. In other words, it cannot be deduced properly from
the second law, if we make no more assumptions than is necessary for
the second law itself.

To show this, we consider a thought experiment with a system called
a {\it xenium engine}. This system undergoes a reversible process in
which the fraction $\delta Q/T$ is not an exact differential. Thus,
the Clausius inequality is violated. However, there is no
contradiction to the second law. There are in fact two xenium
engines, and the violation of the inequality (1) is a consequence of
the entropy exchange between them. The only unusual property of the
process is the absence of any energy transfer or exchange of
particles between this two engines.

At this point the author's intention can easily be misunderstood, so
it is appropriate to give some explanation. What we are going to
prove is that the Clausius inequality (1) cannot be applied wherever
the second law can. So, the xenium engine may be unrealistic and
peculiar, but it is not the point. The engine definitely obeys the
second law, together with a lot of much more exotic systems, real
and imaginary. One cannot possibly expect a perpetuum mobile of the
second kind to be devised by the same means. Thus, this example
shows that the Clausius inequality and the second law is not the
same.

Another thought experiment considered in this paper is the Szilard
engine, invented by L. Szilard in 1929 \cite{S}. Due to its
microscopic nature it is much less convenient for analysis than the
xenium engine, but, in a sense, more realistic. The fact that the
Szilard engine violates the Clausius inequality was obvious since
its inception, for it performs work in an isothermal cycle.
Moreover, the engine converts heat to work, in apparent
contradiction to the second law. The explanation of why the engine
cannot break the second law is known for decades. The explanation,
however, does not rescue the Clausius inequality. This peculiarity
is in essence known to experts but, oddly enough, it has never been
formulated in terms of the inequality, for the best of author's
knowledge.

In the literature, the analysis of the Szilard engine is always
performed by means of statistical mechanics; the only exception is
the work of Ishioka and Fuchikami \cite{IF}, where the Clausius
entropy is considered. Usually, the authors show that what is going
on does not contradict the second law and stop at this point with no
intention to go further. The thermodynamic way of thinking is the
opposite: to take the second law for granted and to find out what
exactly may go on and what may not. This, of course, is a more
difficult problem. The present paper is a modest attempt to address
it.

Recently, the problem of the origin of the Clausius inequality in
statistical mechanics has attracted some attention. The author would
like to stress that in the present paper we follow the thermodynamic
approach exclusively. A seemingly related question about the
validity of the inequality in statistical physics is actually a
different problem. The known proofs of the Clausius inequality by
methods of statistical mechanics (e.g. \cite{J}) have little in
common with the classical argument by Clausius. So, it is by no
means clear what a ``statistical'' analog of the environment
independence condition, which plays a key role in the analysis
below, may be. (The author's guess is that it can only be formulated
for a quantum system). A problem appears to be more difficult in
statistical physics than in classical thermodynamics, which is not
unusual.

The open question is whether the Clausius inequality may be broken
in a real experiment. The inequality is a falsifiable statement, and
one can test it in a laboratory.  Someone who is going to test the
Clausius inequality needs a thermodynamic framework for the
analysis. One of the goals of this paper is to provide such a
framework. One may compare this to the post-Newtonian formalism
which is a tool in tests of general relativity. However, there is
some difference. Unlike the post-Newtonian formalism, the proposed
formalism is in perfect agreement with all the basic principles of
the ``mainstream theory'', i.e. classical thermodynamics. The reason
is simple. Contrary to the common belief, the Clausius inequality is
not a logical equivalent of the second law, but a stronger
statement. So, its violation, if it is really possible, can be
accepted without devastating consequences to the theory. Due to this
subtlety in the internal logic of thermodynamics ``to test the
Clausius inequality'' does not mean ``to try to invent a perpetuum
mobile of the second kind''.

The paper is organized as follows. The next section is devoted to
the xenium engine. In Sec. 3 we discuss the standard proof of the
Clausius inequality. Under close examination, it is based on
implicit assumptions which are wrong for the xenium engine. A
reformulation of the Clausius inequality is proposed in Sec. 4. The
definition of entropy is discussed in Sec. 5. The Szilard engine is
considered in Sec. 6. In Sec. 7 the possibility of experimental test
of the Clausius inequality is discussed. The recent progress in
feedback control \cite{T} makes such a test sufficiently realistic.
The conclusions are summarized in Sec. 8.

\section{THE XENIUM ENGINE}

In this section we consider a kind of heat engine. We call it a {\it
xenium engine}. The working body of the engine is an imaginary gas
with specific properties. We call this gas {\it xenium}, denoted by
the symbol ${\rm Xe }$.

Xenium is an ideal gas. A molecule of xenium is in either of two
states, denoted by ${\rm Xe_a}$ and ${\rm Xe_b}$. The energy levels
of the states are the same. A single molecule can never change its
state. However, two sufficiently close molecules may exchange their
states because of a specific interaction. This may be considered a
sort of chemical reaction:
$${\rm Xe_a}+{\rm Xe_b^{\prime}}
\longleftrightarrow {\rm Xe_b}+{\rm Xe_a^{\prime}};\eqno{(2)}$$
(here ${\rm Xe}$ and ${\rm Xe^{\prime}}$ denote two different
molecules, considered as classical particles). There is a
resemblance to the electron self-exchange, but no particle like
electron is supposed to be transferred. The total number $N_a(N_b)$
of the ${\rm Xe_a}({\rm Xe_b})$ molecules does not change. Thus,
xenium is a mixture of two gases to some extent.

The xenium engine is a cylinder with a piston which moves without
friction (Fig. 1). The wall opposite to the piston is adiabatic. It
is also thin in a sense explained below. The cylinder is divided in
two by a semipermeable partition. The ${\rm Xe_a}$ molecules can
penetrate it while the ${\rm Xe_b}$ ones can not. The space between
the thin wall and the partition, called a {\it camera}, is filled by
a mixture of ${\rm Xe_a}$ and ${\rm Xe_b}$. Another part of the
cylinder is filled with pure ${\rm Xe_a}$.
\begin{figure}
  \centering
  \includegraphics{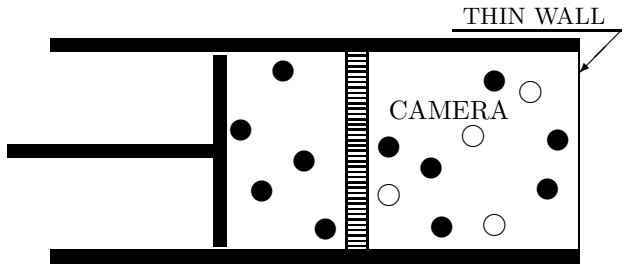}\\
  \caption{The xenium engine}\label{}
\end{figure}

Consider first a single engine in contact with a heat bath at
temperature $T$. The piston moves in a quasistatic (hence
reversible) process. The pressure $P$ on the piston is then equal to
the partial pressure of ${\rm Xe_a}$ in the camera. By the
Gay-Lussac law,
$$P=N_ak_BT/V,$$
where $k_B$ is the Boltzmann constant and $V$ is the volume between
the piston and the thin wall. It is convenient to consider the
dimensionless volume $v=V/V_0$, where $V_0$ is the (constant) volume
of the camera. Work in the process is
$$\delta W=-PdV=-N_ak_BTd\ln v.$$
The internal energy of the gas does not depend on volume, hence
$\delta Q=-\delta W$ and
$$\frac{\delta Q}{T}=N_ak_Bd\ln v.\eqno{(3)}$$

Now consider two xenium engines connected as in Fig. 2. Each engine
is in contact with a particular heat bath. The adiabatic wall
separating the engines is so thin  that xenium molecules in one
camera may interact with molecules in another camera. The
interaction looks similar to a diffusion (Fig. 3).

\begin{figure}
  \centering
  \includegraphics{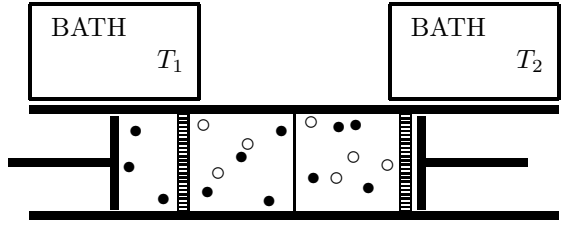}\\
  \caption{Pair of xenium engines}\label{}
\end{figure}

\begin{figure}
  \centering
  \includegraphics{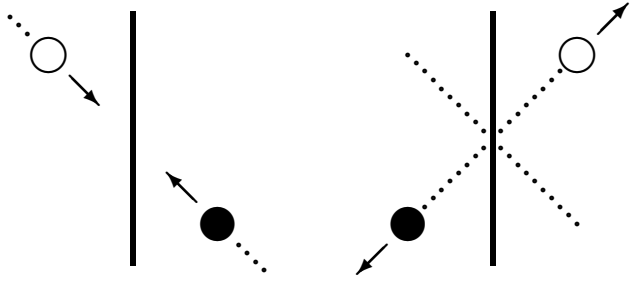}\\
  \caption{Interaction through the thin wall. A molecule of ${\rm Xe_a}$ (black ball)
  turns to ${\rm Xe_b}$ (white ball), and vice versa.}\label{}
\end{figure}

To distinguish variables related to different engines we use
subscripts `1' and `2'. While the total number of ${\rm Xe_a}$
molecules $N_{a,1}+N_{a,2}$ remains a constant, the summands became
functions of two variables $v_1$ and $v_2$. It is not difficult to
find this functions explicitly. Denote by $z$ the quotient of the
concentrations of ${\rm Xe_a}$ and ${\rm Xe_b}$ in the camera:
$$z=\frac{[{\rm Xe_a}]}{[{\rm Xe_b}]}=\frac{N_a}{vN_b}.$$
Then
$$N_a=\frac{N}{1+v^{-1}z^{-1}},$$
where $N=N_a+N_b$ is a constant for each engine.

By symmetry, the equilibrium constant of the reaction (2) is one.
Thus, it comes to equilibrium when $z_1=z_2$. The process is
quasistatic, hence this equality holds all the time. In the simplest
case $N_1=N_2=N_{a,1}+N_{a,2}=N$ we have
$$z_1=z_2=\frac{1}{\sqrt{v_1v_2}}
,\,N_{a,1}=\frac{N\sqrt{v_1}}{\sqrt{v_1}+\sqrt{v_2}},\, \frac{\delta
Q_1}{T_1}= \frac{2k_BN}{\sqrt{v_1}+\sqrt{v_2}}d\sqrt{v_1}.$$
Clearly, $\delta Q_1/T_1$ is {\it not} an exact differential. It is
not difficult to invent a cyclic process such that the Clausius
inequality will be violated for this engine. However, the sum
$$\frac{\delta Q_1}{T_1}+\frac{\delta Q_2}{T_2}=2k_BNd\ln(\sqrt{v_1}+\sqrt{v_2})$$
is an exact differential! (A direct computation shows that it is
true for a general choice of parameters as well).

The conclusions are as follows. The xenium engine undergoes a
reversible process, but the fraction $\delta Q/T$ is not an exact
differential.  Thus, the Clausius inequality can be violated. One
can see that a pair of xenium engines is working as a single Carnot
engine. Thus, no contradiction to the second law may appear. We have
to admit that the Clausius inequality is not a consequence of the
second law alone, without extra assumptions. If it were, a situation
when the second law is valid while the inequality is not, would not
be possible.

In fact, the behavior of the xenium engine can be described by
standard thermodynamics. In this case we have, however, to treat the
engine as if it were not a closed system. The entropy $S$ of the
xenium engine is the sum
$$S=S_a+S_b,$$
where $S_a(S_b)$ is the entropy of ${\rm Xe_a}({\rm Xe_a})$. The
temperature is fixed, and xenium is an ideal gas, hence
$$S_a=-N_ak_B\ln[{\rm Xe_a}],\,S_b=-N_bk_B\ln[{\rm Xe_b}].$$
Taking into account the constrain $dN_b=-dN_a$, we have the
following formula for the change in the entropy of the engine
$$dS=N_ak_Bd\ln v-k_B\ln zdN_a.$$
Thus,
$$dS=\frac{\delta Q}{T}-k_B\ln zdN_a.$$
The ``wrong'' term appears due to the change of composition, which,
from a purely formal point of view, is not possible for a closed
system.

The fundamental equation for the xenium engine is
$$dU=TdS-PdV+(\mu_a-\mu_b)dN_a,$$
where $\mu_a(\mu_b)$ is the chemical potential of ${\rm Xe_a}({\rm
Xe_b})$; note that for ideal gas
$$\mu_a-\mu_b=k_BT\ln[{\rm Xe_a}]-k_BT\ln[{\rm Xe_b}]=k_BT\ln z.$$
The third term on the right hand side is, of course, the chemical
work.

Consider now a pair of xenium engines. We have the constrain
$dN_{a,2}=-dN_{a,1}$, hence the change in the  total entropy due to
the interaction through the thin wall is given by
$$dS_1+dS_2=-k_B(\ln z_1-\ln z_2)dN_{a,1}.$$
The equilibrium condition is then $z_1=z_2$, as expected. One can
see that the ``wrong'' terms in the total entropy are canceled:
$$dS_1+dS_2=\frac{\delta Q_1}{T_1}+\frac{\delta Q_2}{T_2}.$$
Thus, it is clear that the origin of the Clausius inequality
violation is the exchange of entropy between two engines due to the
``chemical'' interaction.

The above entropy calculation leaves one unpleasant question. As is
well known, the very definition of entropy in classical
thermodynamics is based on the Clausius inequality. As the
inequality is not valid, one might ask what exactly the entropy of a
system is and whether it can be defined properly at all. This
reasonable question is answered in Sec. 5.

\section{ENVIRONMENT DEPENDENT PROCESS}

In the following two sections we are going to scrutinize the
foundations of the Clausius inequality. First of all, we need to set
up the terminology. A system in this paper is a closed thermodynamic
system in thermal equilibrium. The latter means that whenever the
system is in contact with a heat bath, it is in thermal equilibrium
with it. Thus, the temperature $T$ in (1) is the temperature of the
system itself as well.  Heat $Q$ is the energy transferred to the
system from a heat bath and work $W$ is the energy transferred to
the system from an external agent. By the first law of
thermodynamics,
$$\Delta U=Q+W,$$
where $U$ is the internal energy of the system. For example, $W=-Q$
in a cyclic process. As usual, we call a process reversible if it is
possible to restore the system as well as the environment to the
initial state.

The question under consideration is if the Clausius inequality (1)
is true for any cyclic process. Many different proofs of this
inequality are known. We have no need to discuss any of these proofs
in detail. They all have essentially the same gap. It is sufficient
to consider a system undergoing an isothermal cyclic process in
contact with a single heat bath. Suppose that the Clausius
inequality is violated, i.e. $Q>0$. Thus, heat is taken from the
bath and, by the first law, converted to work. Is it in
contradiction to the second law? Not yet.

The Kelvin-Planck formulation of the second law states that {\it it
is not possible to take heat  from a single heat bath and convert it
to work in a cyclic process}. The point is, what is a ``cyclic
process'' in this statement. Naturally, {\it any} system is supposed
to undergo a cycle, with the exception of the bath. Feynman \cite{F}
put it as ``{\it a process whose {\bf only} net result is to take
heat from a reservoir and convert it to work is impossible}''. All
the proofs of the Clausius inequality are based on the implicit
assumption that nothing is changed in the environment. Of course,
the inequality $Q>0$ is not possible in this case. The problem is
whether this assumption can be justified or not.

To make the argument clear, let us take the following definition.
Call an {\it adiabatic} process {\it environment independent} if a
system may undergo it causing no change in the environment. An {\it
isothermal} process is environment independent if a system may
undergo it in contact with a single heat bath in such a way that the
couple system+bath undergoes an environment independent adiabatic
process as a whole. A general process is environment independent if
it can be replaced by a combination of isothermal and adiabatic
processes of this kind.

We call a process {\it environment dependent} if it is not
environment independent. It is clear from the definition that if a
system undergoes an isothermal process with this property then
either the environment is changed or heat is given (taken) to the
bath from (by) the environment. (It implies that no system can
undergo a process of this kind when there is this system and a heat
bath and nothing else. This is why the author chose the word
``dependent''). The inequality $Q>0$ for an isothermal cycle does
not contradict the second law in this case, for the best of our
knowledge at least.

The xenium engine considered in Sec. 2 is an example of a system
undergoing an isothermal environment dependent process. Indeed, if
nothing is changed in the environment, then $dN_a=0$, because the
total number of ${\rm Xe_a}$ molecules is unchanged. Thus, any
process with $dN_a\neq 0$ is environment dependent. For such a
process the Clausius inequality is not proved, and is not valid.

\section{THE WEAK CLAUSIUS INEQUALITY}

We have seen that the Clausius inequality cannot be proved for an
environment dependent cycle. Nevertheless, it can be proved for an
environment independent cycle by a slight variation of the standard
method \cite{K}. Consider a system undergoing an environment
independent cyclic process. We can replace it by a combination of
environment independent isothermal and adiabatic processes. Thus,
all the environment remains unchanged, except for a heat bath or
several bathes. It is important that any bath in this process
exchange heat with the system only. Denote by $Q_j$ the heat taken
by the system from the bath at temperature $T_j$. Then, the heat
taken by this bath is $-Q_j$.

Let us introduce one more heat bath at temperature $T_0$. With the
help of the Carnot engine ($=$ environment independent reversible
cyclic device) we can restore every bath, save this one, to its
initial state by giving it heat $Q_j$ at the expense of heat taken
from the exceptional bath at temperature $T_0$. The net result of
the process will be to take heat $Q_0$ from this bath and convert it
to work. By the second law, $Q_0\le 0$. Taking into account the
properties of the Carnot engine, we have the equality
$$\sum_j\frac{Q_j}{T_j}=\frac{Q_0}{T_0},$$
and the Clausius inequality follows.
Going to the limit, we can replace the sum by the integral and write
it in the common form (1).

The argument fails for a system undergoing an environment dependent
process because of the change in other systems which should be taken
into account. Consider a number of systems undergoing a cyclic
process together. Following the classical scheme, we connect all the
systems to the same bath at temperature $T_0$ through the Carnot
engines (Fig. 4). We suppose that any system involved in the process
is taken into account, hence the rest of the environment remains
unchanged. Then, by the second law, $Q_0\le 0$, where $Q_0$ is the
heat taken from the bath. However, the quotient $Q_0/T_0$ is now
equal not to a single integral but to the sum of integrals taken
over all the systems. Thus, the inequality we have is the following
$$\sum_i \oint\frac{\delta Q_i}{T_i}\le 0,\eqno{(5)}$$
where the subscript $i$ is related to $i$-th  system.
Call this a {\it weak Clausius inequality}.

\begin{figure}
  \centering
  \includegraphics{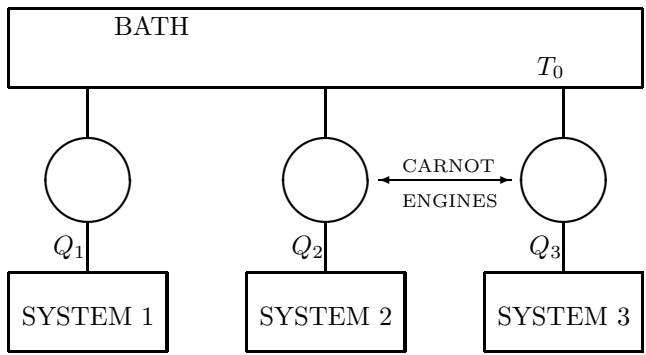}\\
  \caption{Three systems connected to a heat bath.}\label{}
\end{figure}

In the case of a reversible process (5) turns to an equality. By the
common argument, there exists a function $S$ such that
$$dS=\sum_i \frac{\delta Q_i}{T_i}.$$
We have seen an example in Sec. 2. Obviously, $S$ is the total
entropy. The definition of the entropy of an individual system is
considered in the next section.

\section{ENTROPY}

In classical thermodynamics the definition of entropy is grounded on
the Clausius inequality. What happens to entropy if the inequality
fails? The answer to this question become obvious as soon as we
realize that any process in the classical setting is implicitly
assumed to be environment independent. All we have to do is to make
this assumption explicit. Thus, the entropy of a system should be
defined by the familiar formula
$$S_B-S_A=\int_{A}^B\frac{\delta Q}{T},$$
where $S_A(S_B)$ denotes the entropy of the system in the state
$A(B)$, and the integral is taken over an arbitrary {\it environment
independent} reversible process $A\to B$. From the previous section
we know that for an environment independent cycle the Clausius
inequality is valid, hence the definition is sound. Of course, we
have to assume that any two states may be connected by an
environment independent process (which is not obvious sometimes).

By definition, the change in entropy in a reversible environment
independent process is $dS=\delta Q/T$, but for a general reversible
process this equality may be wrong. Denote the difference by
$$\delta{\mathfrak S}=dS-\frac{\delta Q}{T}.\eqno{(6)}$$
Call $\delta{\mathfrak S}$ the (infinitesimal) {\it adiabatic
entropy}, taken by the system. Note that by the weak Clausius
inequality
$$\sum_i {\mathfrak S}_i=0,$$
where the sum is taken over all the systems involved in the process.
Thus, adiabatic entropy is a form of entropy transfer, by the same
way as work is a form of energy transfer.

If the process is neither environment independent, nor reversible,
then the equality (6) turns to an inequality. Taking the integral
over a cycle, we get the following generalization of the Clausius
inequality
$$\oint\frac{\delta Q}{T}\le -{\mathfrak S}.\eqno{(7)}$$
(Here is a subtle point. To define adiabatic entropy taken in an
irreversible process we have to assume that all the environment
undergoes a reversible process. The adiabatic entropy taken by a
system is then the adiabatic entropy given by the environment).

In Sec. 2 the entropy of the xenium engine was calculated by
standard thermodynamic rules. Here we present a formal proof that it
is the ``right'' entropy. (Apparently, there is no real need of such
a proof. It is given by way of  illustration of the argument). First
of all we have to make sure that any process under consideration is
environment independent. For this reason we consider a {\it single}
xenium engine attached to a heat bath. This system has two
parameters: $v$ and $N_a$. The simplest process is the piston
moving. In this case we have, by (3),
$$dS=N_ak_Bd\ln v,\,N_a={\rm const}.$$

Then, we run into an obstacle: in our model the parameter $N_a$ is a
constant in any environment independent process, so we cannot
measure the entropy difference between the states with different
numbers $N_a$. To circumvent this obstacle we have to make the model
a bit more realistic. Let us suppose that there  exists a catalyst
which makes xenium molecules undergo spontaneous transitions ${\rm
Xe_a}\leftrightarrow{\rm Xe_b}$. Consider the following process. At
the beginning $N_a=N/2$ and $v=1$. We add catalyst into the camera
and move the piston out. Due to the catalyst, we have $z=1$ all the
time, hence $N_a=N/(1+v^{-1})$. The entropy change, of course, is
given by the same formula $dS=N_ak_Bd\ln v$.

This is enough to find the entropy in the case $N_a\le N/2$, but the
direct computation is a bit ugly.(To access states with $N_a>N/2$ we
need a more complicated process). Instead, one may notice that for
the both processes we have the equality
$$dS=N_ak_Bd\ln v-k_B\ln zdN_a;$$
the second term on the right hand side vanishes either because of
$dN_a=0$, or because of $\ln z=0$. But there are exact differentials
on the both sides, hence the equality remains true in general. As
one might expect, the adiabatic entropy taken by the xenium engine
is
$$\delta{\mathfrak S}=\frac{\partial S}{\partial N_a}dN_a,$$
that is, it is the entropy gained due to the change of composition.

The work done on a system in a reversible isothermal cyclic process
can be found from (7):
$$W=T{\mathfrak S}.$$
So, the total work performed by a pair of xenium engines is zero if
$T_1=T_2$. In this case the process can be described by the diagram
$${\rm heat}\longrightarrow {\rm work}+{\rm adiabatic\, entropy}
\longrightarrow{\rm heat}$$ One of the engines takes heat from a
bath, converts it to work and gives adiabatic entropy to another
engine. The latter converts work back to heat and gives heat to
another bath.

\section{THE SZILARD ENGINE}

 There is a branch of thermodynamics where the Clausius inequality
violation cannot be ignored. It is thermodynamics of a feedback
controlled system or, to be more precise, a branch of thermodynamics
which deals with problems commonly related to the famous Maxwell's
demon \cite{M}. A common property of a feedback controlled system is
the ability to perform work in an isothermal cycle. Such a behavior
formally contradicts both the Clausius inequality and the second
law. There exists a well known explanation of why the contradiction
to the second law is not real. But, under close examination, the
Clausius inequality is broken indeed.

Here we consider the Szilard engine which is, so to say, the
prototype of a feedback controlled system. This imaginary device was
invented by L. Szilard in 1929 \cite{S}. It consists of a box with a
single particle. The box is provided with a thin piston which can be
inserted to or removed from it as necessary (Fig 5). The engine
undergoes an isothermal cyclic process in contact with a heat bath
at temperature $T$. At the beginning, the piston is out of the box.
As the first step of the process it is inserted into the box at the
middle, dividing it into two parts of equal volume. The particle
gets trapped in one of the halves. After that, the piston moves into
the empty half until it reaches the wall. The piston is then removed
and the cycle is complete.

\begin{figure}
  \centering
  \includegraphics{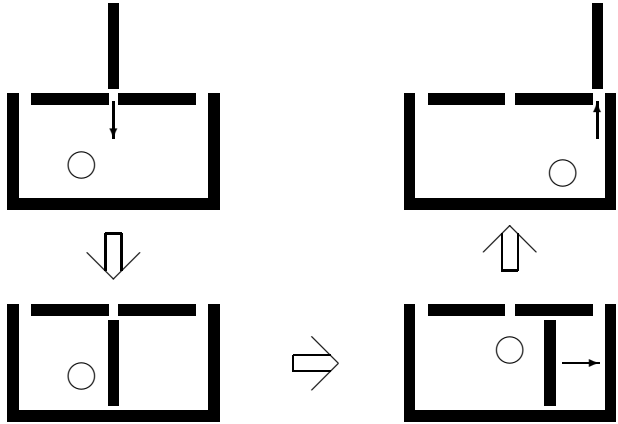}\\
  \caption{The Szilard engine}\label{}
\end{figure}

It is supposed that the particle termalizes in any collision with
the walls. It is also supposed that the single-particle gas expands
reversibly. Under this assumptions the Gay-Lussac law is valid:
$P=k_BT/V$. Thus, the work performed in the cycle on the system is
$$W=-\int_{V/2}^{V} PdV=-k_BT\ln 2.$$
The problem considered in the literature is why the engine cannot
violate the second law. We consider a different problem: why the
engine {\it can} violate the Clausius inequality. There exists a
quite extensive literature on the Szilard engine \cite{B,Fa,M,P}, so
we have no need to discuss numerous technicalities related to this
complicated subject.

It is well known that peculiar thermodynamic properties of the
Szilard engine are related to the fact that it cannot work by
itself. It needs a {\it controller}. This is a device which makes a
{\it measurement} to find out in which half of the box the particle
gets trapped and drives the engine. The Szilard engine cannot work
in the absence of a controller. Once a controller is taken into
account properly, the paradox with the apparent violation of the
second law dissolves. For the best of author's knowledge, this was
first noticed by C.H.Bennett \cite{B}; now it is a commonplace.

The presence of a controller, however, does not alter the fact that
the Szilard engine performs work in an isothermal cycle. Thus, while
the second law is beyond doubt, a paradox remains. Actually, it has
been discussed in the literature \cite{P,IF,M}, but the discussion
was mostly limited to statistical mechanics. The Clausius inequality
was virtually ignored, and the problem was interpreted as a
misbehavior of the Boltzmann-Gibbs entropy. (Not the Clausius
entropy. For a realistic system it is essentially the same entropy,
but an approach makes a difference).

Apparently, the only work where the  Clausius, i.e. thermodynamic
entropy of the Szilard engine is considered, is the paper of Ishioka
and Fuchikami \cite{IF}. In their paper, however, the Clausius
inequality is not mentioned\footnote{The same authors did mention
the Clausius inequality in the preprint \cite{IF0}, and admitted
that it is {\it valid}, though not with confidence. (Which may be
considered a forgivable mistake). However, in a subsequent paper
\cite{IF}, which provides a much more detailed account of the same
subject, there is no single word about the inequality. This makes
the author quite sure that the provocative question if the
inequality is true for the Szilard engine was avoided by
intention.}.
 According to \cite{IF}, the Clausius entropy of the engine
decreases by $k_B\ln 2$ when the piston is inserted into the box
\footnote{The same is true for the Boltzmann-Gibbs entropy
\cite{P}}. This effect cannot be explained in terms of the usual
classical thermodynamics, for a very simple reason. The insertion is
not a process, it is in fact two different processes, with the same
initial state but with different final states. The point is, the
choice between the processes is random. A ``macroscopic randomness''
\cite{P} of this kind is beyond the scope of conventional classical
thermodynamics, which is based entirely on the determinism.

The author's suggestion is to extend slightly the framework of
thermodynamics by introducing the following variant of the Clausius
equality

$$\oint_{\mathcal{A}}\frac{\delta Q}{T}=
k_B\ln\frac{P(\mathcal{A}^{-1})}{P(\mathcal{A})},\eqno{(8)}$$

where $\mathcal{A}$ is a cyclic process, reversible in a sense. Here
$P(\mathcal{A})$ denotes the probability of $\mathcal{A}$, i.e. the
probability of success in an attempt to make a system undergo this
process. The equality is certainly true for the Szilard engine and
the generalizations (like a ``skewed'' engine \cite{M}), but its
range of validity in general is not certain. Under appropriate
assumptions it can be proved by means of statistical mechanics, by
phase space calculus or following the lines of \cite {SU2}, but the
author would like to focus attention on the thermodynamic side of
the picture. In classical thermodynamics, given the state of the
art, the equality (8) may only be regarded as a plausible
conjecture. The author prefers to consider it a possible formulation
of the so-called Landauer's principle \cite{L}.

A violation of the Clausius inequality by the Szilard engine can,
and should, be explained at two different levels. At the low level,
it is a consequence of ``macroscopic randomness''. Depending on the
location of the particle after the insertion, the engine undergoes
one of two processes with equal probability,
$P(\mathcal{A}_1)=P(\mathcal{A}_2)=1/2$. The reverse processes are
both deterministic, $P(\mathcal{A}_1^{-1})=P(\mathcal{A}_2^{-1})=1$.
By (8), $Q=k_BT\ln 2$ in any case.

At the high level, it is a consequence of environment dependence.
(Note that, strictly speaking, we cannot apply this concept to a
process which is not deterministic. To circumvent this obstacle we
have to consider the ``probabilistic'' part of the process, from the
insertion of the piston to the memory erasure in a controller, as a
whole, without separating it into stages.) The very fact that the
engine does not work properly in the absence of a controller implies
that it undergoes an environment dependent process. For this reason,
the fact that the engine performs work in an isothermal cycle does
not contradict anything. Following the same protocol as in the case
of the xenium engine, we can attach the Szilard engine to a heat
bath and the controller to another heat bath. The standard analysis
\cite{B,Fa,P,IF} then shows that the entropy of the former bath
decreases by $k_B\ln 2$ per cycle while the entropy of the latter
one increases by the same amount. There is no heat exchange between
the engine and the controller, hence there is adiabatic entropy
transfer between these systems.

\section{TESTING THE VIOLATION OF THE CLAUSIUS INEQUALITY}

In this section we discuss briefly the possibility of experimental
detection of the violation of the Clausius inequality. For obvious
reasons we may restrict ourselves by isothermal processes.
 For an isothermal cycle the Clausius inequality  is equivalent
to the work inequality $W\ge 0$. A violation of the latter
inequality for the {\it total} work is forbidden by the second law
of thermodynamics. However, we may consider a process which involves
several interacting systems. If $W_i$ denotes the work done on  i-th
system, then, by the second law,
$$\sum_i W_i\ge 0.$$
On the other hand, by the Clausius inequality, $W_i\ge 0$ for {\it
each} system. The latter is a more strong statement than the former.
Due to this difference, a violation of the Clausius inequality under
appropriate conditions does not imply an opportunity for  building a
perpetuum mobile of the second kind.

A promising approach to test the Clausius inequality is to perform
an experiment with feedback control. All we need here are some basic
principles. We follow the convenient notation of \cite{SU}. There
are three thermodynamic systems taking part in the process: a
controlled system S, a memory (or controller) M, and a heat bath B.
Let $W^S$ be the work done on S and $W^M$ be the work done on M. For
an isothermal cyclic process we have two inequalities:
$$W^M\ge k_BTI, W^S\ge -k_BTI,$$
where $T$ is the temperature of B and $I\ge 0$ is the mutual
information. The quantity $I$ may be interpreted as the amount of
information about S, obtained during the cycle. It follows that
$W^S+W^M\ge 0$, in agreement with the second law.

In the process considered in \cite{SU} no violation of the Clausius
inequality is possible, because the cycle is not actually complete
(the state of S is changed). To make the process a proper cycle we
must extend it. Let S be initially in thermodynamic equilibrium in
contact with B. Then it is detached from B and the process  goes on
as in \cite{SU}. Finally, S is attached to B again. We have then an
isothermal cyclic process. If $I>0$  then the possibility of the
following inequality
$$-k_BTI\le W^S<0$$
is not excluded by any known principle. This possibility, if
realized, would imply a violation of the Clausius inequality by S.
This consideration is in fact very general and independent of the
details of the process. To test the Clausius inequality we have
simply to measure the work $W^S$.

A successful feedback control experiment has been recently performed
\cite{T}. In the experiment, a small particle is rotated against the
applied moment of a force, at the expense of heat taken from the
environment. Though this experiment was not designed to test the
Clausius inequality, the author is at the opinion that the
possibility of a violation of the inequality under similar
conditions ought to be discussed. The violation, of course, is by no
means obvious. One of the major issues is the influence of the
measurement on the process. (The particle is illuminated to find out
its position. The light is absorbed partially by the particle and
the medium, hence the process is not exactly isothermal).

An interesting question is if the Clausius inequality can be
violated on macroscopic scale. Quantitatively, the question is about
the possibility of ``large scale'' adiabatic entropy transfer
${\mathfrak S}/k_B \gg 1$. At present, the author does not know how
such an experiment can be devised. On the other hand, he does not
know about any fundamental obstacle either.

\section{CONCLUSIONS}

In this paper the limitations of the Clausius inequality are
discussed. It is shown that in the general case the inequality is
not a consequence of first principles of thermodynamics. Thus, the
Clausius inequality violation is not forbidden, for the best of our
knowledge.

Such a violation, if it is possible, can be explained in terms of
thermodynamics by adiabatic entropy transfer between two systems. In
this hypothetical process entropy is transferred from one system to
another while there is neither energy transfer nor exchange of
particles. This transfer  takes place in some thought experiments,
including the well known Szilard engine.

The possibility of the Clausius inequality violation in nature is
not excluded by any known principle of physics. So, it would not be
unreasonable to test it in a laboratory. The recent progress in
feedback control experiments makes such a test realistic enough.

{Alexey~V.~Gavrilov, Department of Physics, Novosibirsk State
University, Russia}
\begin{flushright}
\email{gavrilov19@gmail.com}
\end{flushright}

\end{document}